\documentclass[twocolumn,aps,prb,superscriptaddress,a4paper,showpacs,showkeys,lengthcheck]{revtex4-1}

\usepackage{graphicx}
\usepackage{dcolumn}
\usepackage{bm}
\usepackage{type1cm}

\usepackage{amssymb,amsmath}

\usepackage[dvips]{color}

\newcommand{\mm}[1]{\mbox{$#1$}}

\newcommand{\ms}{\mbox{$\mu_{\mathrm{spin}}$}}
\newcommand{\mo}{\mbox{$\mu_{\mathrm{orb}}$}}
\newcommand{\moms}{\mbox{$\mu_{\mathrm{orb}}/\mu_{\mathrm{spin}}$}}

\newcommand{\momsd}{\mbox{$\mu_{\mathrm{orb}}^{(d)}/\mu_{\mathrm{spin}}^{(d)}$}}
\newcommand{\mB}{\mbox{$\mu_{B}$}}

\newcommand{\Led}{{\it L$_{2,3}$}\ edge}
\newcommand{\Leds}{{\it L$_{2,3}$}\ edges}

\newcommand{\ldau}{LSDA+$U$}
\newcommand{\dmft}{LSDA+DMFT}

\newcommand{\ea}{{\it et al.}}

\begin{document}

\title{Many-body effects in x-ray absorption and magnetic
    circular dichroism spectra within the LSDA+DMFT framework}



\author{O. \surname{\v{S}ipr}} 
\email{sipr@fzu.cz}
\homepage{http://www.fzu.cz/~sipr} 
\affiliation{Institute of Physics of the ASCR v.~v.~i.,
  Cukrovarnick\'{a}~10, CZ-162~53~Prague, Czech Republic }

\author{J. \surname{Min\'{a}r}}
\affiliation{Universit\"{a}t M\"{u}nchen, Department Chemie und
  Biochemie, Butenandtstr.~5-13, D-81377~M\"{u}nchen, Germany}

\author{A. \surname{Scherz}} 
\affiliation{Stanford Institute for Material and Energy Science, SLAC
  National Accelerator Laboratory, Menlo Park, California, USA}

\author{H. \surname{Wende}} 
\affiliation{Universit\"{a}t Duisburg-Essen, Fakult\"{a}t f\"{u}r
  Physik and Center for Nanointegration Duisburg-Essen (CeNIDE),
  Lotharstr.~1, 47048~Duisburg, Germany} 

\author{H. \surname{Ebert}}
\email{Hubert.Ebert@cup.uni-muenchen.de}
\homepage{http://ebert.cup.uni-muenchen.de}
\affiliation{Universit\"{a}t M\"{u}nchen, Department Chemie und
  Biochemie, Butenandtstr.~5-13, D-81377~M\"{u}nchen, Germany}

\date{\today}

\begin{abstract}
The theoretical description of photoemission spectra of transition
metals was greatly improved recently by accounting for the
correlations between the $d$\ electrons within the local spin density
approximation (LSDA) plus dynamical mean field theory (DMFT).  We
assess the improvement of the LSDA+DMFT over the plain LSDA in x-ray
absorption spectroscopy, which --- unlike the photoemission
spectroscopy --- is probing unocccupied electronic states.  By
investigating the \Led\ x-ray absorption near-edge structure (XANES)
and x-ray magnetic circular dichroism (XMCD) of Fe, Co, and Ni, we
find that the \dmft\ improves the LSDA results, in particular
concerning the asymmetry of the $L_{3}$\ white line.  Differences with
respect to the experiment, nevertheless, remain --- particularly
concerning the ratio of the intensities of the $L_{3}$\ and
$L_{2}$\ peaks.  The changes in the XMCD peak intensities invoked by
the use of the \dmft\ are a consequence of the improved description of
the orbital polarization and are consistent with the XMCD sum rules.
Accounting for the core hole within the final state approximation does
not generally improve the results.  This indicates that to get more
accurate \Led\ XANES and XMCD spectra, one has to treat the core hole
beyond the final state approximation.
\end{abstract}



\maketitle


\section{Introduction}   \label{sec-intro}

X-ray absorption spectroscopy (XAS) evolved into a powerful technique
for studying electronic as well as geometric structure of solids.  Its
main strength include chemical selectivity, angular-momentum
selectivity and ability to provide detectable signals even for low
amounts of material. This makes it well-suited for studying defects,
adsorbates, or nanostructures.  For studying magnetism, x-ray magnetic
circular dichroism (XMCD) spectroscopy, based on exploring the
energy-dependence of the difference in the absorption of left- and
right-circularly polarized x-rays in a magnetized sample, proved to be
a very powerful tool.\cite{Wen04}

An efficient use of x-ray absorption spectroscopy requires a
significant input from theory.  {\em Ab initio} calculations of x-ray
absorption near-edge structure (XANES) and XMCD are usually quite
successful in reproducing the positions of spectral peaks, fairly
successful in reproducing their intensities and less successful in
reproducing detailed shapes of the peaks.  The severity of the
failures of the theory varies depending on what material is studied,
which absorption edges are involved and what purpose the spectroscopic
measurement serves.

Magnetic 3$d$\ elements are often used in artificial structures to
form materials with properties that are interesting both fundamentally
and for their possible technological application. XAS at the \Leds\ of
Fe, Co, and Ni is used to get information about electronic states of
the $d$\ character, which dominate close to the Fermi energy $E_{F}$. {\em
  Ab initio} calculations based on the local spin density
approximation (LSDA) to density functional theory suffer here from
some common deficiencies.  One such deficiency is the inability to
reproduce correctly the ratio of the intensities of the $L_{3}$\ and
$L_{2}$\ white lines in the XANES.  This has been ascribed to the lack
of proper dynamic treatment of the core hole.\cite{SE98,ANR03} Other
common deficiencies of {\em ab-initio} calculations include lack of
asymmetry of the theoretical XANES white line and underestimated the
ratio of the $L_{3}$\ and $L_{2}$\ XMCD peak
intensities.\cite{AWW98,Wen04,SE05} Having an {\em ab-initio} method
able to deliver a more accurate quantitative agreement with experiment
would be a great help when dealing with complex systems.  The
procedurally simple XMCD sum rules\cite{CTAW93,TCSvdL92} proved to be
very powerful in interpreting experiments but their use has
limitations.  A more robust way is to compare measured spectra to
spectra of a well-defined reference material.  The properties of the
reference material may, nevertheless, differ from the properties of
the investigated system and, moreover, a suitable reference may not be
available.  In such situations, a comparison with accurate and
reliable calculations may be very useful.\cite{SWB+02,Bab05,WSS+07}

Even though the 3$d$\ transition metal (TM) elements can be seen as
moderately correlated materials, there are known effects where
including correlations is necessary.  E.g., the orbital magnetic
moment \mo\ is underestimated by the LSDA and improved results can be
obtained by accounting for the enhancement of the orbital polarization
either via the scheme of Brooks (OP Brooks)\cite{EJA+90} or via the
\ldau\ scheme.\cite{Sol05,SLT98} These schemes, however, account only
for static effects of the electron self-energy.  To describe the
spectra, dynamical effects should be included as well.  This can be
achieved via the LSDA plus dynamical mean field theory (DMFT)
scheme. The \dmft\ formalism proved to be rather successful when
dealing with the photoemission spectra of 3$d$\ transition metals
(TMs).\cite{MEN+05,BME+06,SFB+09,SMB+10} One can, therefore, expect
that this formalism might lead to a substantial improvement also for
XANES and XMCD spectra.  In particular, as the \dmft\ method provides
correct values for \mo,\cite{CMK08,SMME08} one can presuppose that it
should also lead to a better ratio of the $L_{3}$\ and $L_{2}$\ XMCD
peak intensities because these are, within certain considerations,
related to \mo.\cite{CTAW93}

This paper reports on an application of the \dmft\ formalism to x-ray
absorption spectroscopy focusing on the \Led\ spectra of Fe, Co, and
Ni.  We demonstrate that accounting for the valence-band and
conduction-band correlations via the \dmft\ improves the calculated
spectra with respect to the LSDA.  However, the improvement still does
not lead to a fully satisfying reproduction of the experimental data
--- not even if the core hole is included via the final state
approximation.  Based on these results, we conclude that dealing with
the dynamical aspects of the correlations between the core hole and
the valence and conduction electrons is needed for further
investigations.


\section{Computational scheme}   \label{sec-comput}

A detailed description of the implementation of the \dmft\ within the
Korringa-Kohn-Rostoker (KKR) band structure scheme can be found in
previous publications.\cite{MCP+05,Min10} Accordingly, we summarize
here just the major features.  The \dmft\ method belongs to
Hubbard-$U$ band-structure schemes, i.e., the one-electron LSDA
Hamiltonian is extended by an additional Hubbard-Hamiltonian term
which explicitly describes the on-site interaction between (in our
case) the $d$ electrons.  The many-body Hamiltonian is specified by
parameters representing the Coulomb matrix elements.

The main idea of the DMFT is to map the periodic many-body problem
onto an effective single-impurity problem that has to be solved self
consistently.  For this purpose one describes the electronic
properties of the system in terms of the single particle Green's
function $\hat G(E)$, which is determined by
\begin {equation}
\label{dyson}
\left[ E-\hat H_{\mathrm{LSDA}}-\hat \Sigma(E) \right] \hat G = \hat 1 \;\;,
\end {equation}
where $E$ is the complex energy, $\hat H_{\mathrm{LSDA}}$\ is the LSDA
Hamiltonian and $\hat \Sigma$ is a single-site effective self-energy
operator.  Within the DMFT, the self-energy $\hat \Sigma(E)$ is a
solution of the many-body problem of an impurity placed in an
effective medium. This medium is described by the so called {\em bath}
Green's function $\hat{\mathcal G}$ connected to the Green's function
$\hat G(E)$ by
\begin {equation}
\label {bath}
\mathcal{\hat{G}}^{-1}(E) = 
\hat{G}^{-1}(E) + \hat{\Sigma}(E) \;\;.
\end {equation}
For a more detailed description of the DMFT equations the authors
redirect the reader to one of the excellent
reviews.\cite{KSH+06,Hel07}

The self-energy $\hat \Sigma(E)$ and the bath Green's function
$\mathcal{\hat{G}}(E)$ have to be determined self consistently.
Technically, this is done in two steps.  The first step is solving
Eq.~(\ref{dyson}) by the means of spin-polarized fully relativistic
KKR band structure method.\cite{Ebe00} We relied on the atomic sphere
approximation (ASA) to the potential and used an angular momentum
cut-off \mm{\ell_{\mathrm{max}}}=3.  The integration over the
$\mathbf{k}$ points was done on a regular mesh, using 2600 points in
the irreducible part of the Brillouin zone in the case of bcc Fe and
fcc Ni and 900 points in the case of hcp Co.  The Vosko, Wilk, and
Nusair parametrization for the local exchange and correlation
potential was used.\cite{VWN80}

In the second step of the \dmft\ calculation, the self-energy $\hat
\Sigma(E)$ has to be found according to Eq.~(\ref{bath}).  This is
done by solving the many-body effective impurity problem, often
referred to as the DMFT solver.\cite{KSH+06,Hel07} We used
perturbative solvers, either the spin-polarized $T$-matrix + FLEX
solver\cite{PKL05} or the spin-polarized $T$-matrix approximation
solver (TMA).\cite{Cha07D} The use of perturbative solvers is
justified because the correlation effects in pure 3$d$ TMs are not
very pronounced.  The results are very similar for both
solvers. Unless explicitly stated otherwise, data for the TMA solver
are shown here.  For the intra-atomic Hund exchange interaction
$J$\ we take a common value of $J$=0.9~eV.\cite{AG91,Hel07} The
screened on-site Coulomb interaction $U$ is set to 1.7~eV for Fe,
2.3~eV for Co, and 2.8~eV for Ni.  These values were chosen because
they lead to good values of \ms\ and \mo\ (see Chadov \ea\cite{CMK08}
for an extensive study) and also to a correct description of angular
resolved photoemission spectra.\cite{BME+06,SFB+09,SMB+10} Similar
values of $U$ and $J$ were used also by other authors dealing with
these systems,\cite{YSK01,GDK+07,MF08} even though it should be noted
that the parameters $U$ and $J$ are not directly transferable from one
work to another because they depend, among others, on the choice of
the basis set.\cite{Hel07}

Our implementation of the \dmft\ method is self-consistent not only in
the self-energy \mm{\hat \Sigma(E)}\ but also in the charge density
$\rho({\mathbf{r}})$, i.e., in each iteration a new potential is used
to generate a new single particle Green's function $\hat
G(E)$\ entering Eq.~(\ref{bath}). During the self-consistency cycle,
the self-energy \mm{\hat \Sigma(E)}\ is calculated either on a set of
Matsubara frequencies (FLEX solver) or on the real energy axis (TMA
solver).  When calculating x-ray absorption spectra, the self-energy
on the real axis above $E_{F}$ is obtained via the Pad\'{e} analytic
continuation.\cite{VS77}

The LSDA accounts already to some extent for the correlation of the
$d$ electrons, so a corresponding term has to be subtracted, to avoid
counting this interaction twice.  The exact form of this ``double
counting term'' is not available because the LSDA is not formulated in
a diagrammatic language; the choice has to be made by an educated
guess.  One way of doing this is to make {\em a priori} assumptions
about the occupation of the correlated orbitals.  In the limit of a
uniform occupancy of these orbitals, the energy correction to the LSDA
is due to ``fluctuations'' away from the spin-dependent orbitally
averaged occupation.  The interaction term in this around mean-field
(AMF) limit is\cite{CS94}
\begin{eqnarray}
V^{\mathrm{LSDA+AMF}}_{m \sigma} & = & \sum_{m^{'}}   
U_{m m^{'}} ( n_{m^{'} -\sigma} - n^{0}_{-\sigma} ) 
 \\
  &  & +
\sum_{m^{'}\ne m}   
(U_{m m^{'}} - J_{m m^{'}}) ( n_{m^{'} \sigma} - n^{0}_{\sigma} ) 
\;\; .  \nonumber
\end{eqnarray}
In the above equation, $n_{m\sigma}$ is the occupation number for
electrons with orbital and spin quantum numbers $m$\ and $\sigma$,
$n^{0}_{\sigma}$ is the orbitally averaged occupation number and
$U_{mm^{'}}$ and $J_{mm^{'}}$ are matrix elements defined by the
parameters $U$\ and $J$.  The opposite limiting case concerning the
occupation of the orbitals is the around atomic limit (AAL), which
produces the correct behavior if $n_{m\sigma}$=0 or 1. It is sometimes
referred to as the fully localized limit (FLL) and the corresponding
interaction term is\cite{CS94}
\begin{equation}
V^{\mathrm{LSDA+AAL}}_{m \sigma} = V^{\mathrm{LSDA+AMF}}_{m \sigma} -
( U - J ) ( n^{0}_{\sigma} - \frac{1}{2} ) \;\; . 
\end{equation}
The results of the \ldau\ or \dmft\ calculations may strongly depend
on the choice of the double counting (d.c.) procedure.  Based on
earlier works, especially in the field of photoemission, it appears
that the AMF recipe is the most appropriate for the 3$d$
TMs. \cite{SFB+09,SMB+10} It is interesting to see whether this
recipe works also for x-ray absorption spectroscopy (XAS).

We investigate also the effect of the core hole treated within the
final state approximation as fully relaxed and screened, i.e., with
the potential calculated with one electron transferred from the
2$p$\ core level into the valence band.  This was achieved by first
performing a self-consistent calculation without the core hole and
then by treating the photoabsorbing atom with the core hole as a
perturbation, employing the impurity cluster Green's function
method.\cite{BZLD84,Zel88,JR10} In this impurity calculation the
electronic structure was allowed to relax within the first two
nearest-neighbor atomic shells around the photoabsorbing atom.

The theoretical spectra were broadened to account for the finite
lifetimes of the core hole and of the photoelectron.  The core hole
related broadening was simulated by a Lorentzian with full width at
half maxima of 0.40~eV at the $L_{3}$\ edge and 0.70~eV at the
$L_{2}$\ edge, which is in the range of generally accepted
values.\cite{SAB+92,CP01} The excited photoelectron related broadening
was simulated by a Lorentzian with energy-dependent width, according
to the ``universal curve'' as suggested by M\"{u}ller \ea\cite{MJW82}
By optimizing the broadening along the suggestion of Benfatto
\ea\cite{BDN+03} a better visual agreement of our calculations with
experiment could be achieved.  However, this would have no influence
on the conclusions.

Intuitive understanding as well as quantitative analysis of XMCD
spectra has been greatly helped by the XMCD sum rules.  These rules
associate areas of XANES and XMCD peaks with \ms\ and \mo\ of the
photoabsorbing atom.  Our calculations provide \ms\ and \mo\ as well
as XANES and XMCD spectra.  Accordingly, application of the sum rules
to our calculated spectra makes it possible to assess to what extent
the changes in the spectra caused by including the valence- and
conduction-band correlations via the \dmft\ are consistent with the
corresponding changes in the magnetic moments.

For the \Led\ spectra the sum rules can
be written as \cite{Sto99,CTAW93,TCSvdL92}
\begin{equation}
 \frac{3}{I} \, \int \left( \Delta \mu_{{L}_3}
        -2\Delta \mu_{{L}_2} \right) \, \mathrm{d}E  \, = \, 
        \frac{\mu_{\mathrm{spin}}^{(d)} + 7 T_{z}^{(d)}}{n_{h}^{(d)}}
\label{eq-spin}
\end{equation}
and
\begin{equation}
\frac{2}{I}
\int \left(\Delta \mu_{{L}_3}
         + \Delta \mu_{{L}_2} \right)\, \mathrm{d}E \, = \, 
         \frac{\mu_{\mathrm{orb}}^{(d)}}{n_{h}^{(d)}}  \;\; ,
\label{eq-orb}
\end{equation}
where \mm{\Delta \mu_{{L}_{2,3}}} are the differences
\mm{\Delta\mu=\mu^{(+)}-\mu^{(-)}} between the absorption coefficients
for the left and right circularly polarized light at the $L_{2}$ and
$L_{3}$ edges, $I$\ is the integrated isotropic absorption spectrum,
$\mu_{\mathrm{spin}}^{(d)}$\ and $\mu_{\mathrm{orb}}^{(d)}$\ are the
$d$\ components of the local spin and orbital magnetic moments,
$n_{h}^{(d)}$\ is the number of holes in the $d$\ band and
$T_{z}^{(d)}$\ is the $d$\ component of the intra-atomic magnetic
dipole operator for spin quantization axis aligned along $z$.  The
$T_{z}^{(d)}$\ term is negligible for high-symmetry systems as those
dealt with here. However, this does not necessarily apply to more
complex systems.\cite{SME09}
Application of the sum rules (\ref{eq-spin})--(\ref{eq-orb}) requires
setting the energy cut-off $E_{C}$\ which defines the upper boundary
of the 3$d$\ band.  We determined it by requiring that the integrated
density of the $d$\ states is 10 when integrated from the bottom of
the valence band up to $E_{C}$, similar as in our earlier
study.\cite{SME09}


\section{Experiment}   \label{sec-exp}

The bulk-like Fe, Co, and Ni films were grown {\em in situ} directly
at the synchrotron radiation facility BESSY II (Berlin, Germany). As
the substrate a Cu(100) single crystal was used and the films were
prepared at room temperature in ultra high vacuum conditions (base
pressure $2 \times 10^{-10}$ mbar) by evaporation from high purity
rods using a commercial triple e$^-$-beam evaporator. The surface of
the Cu crystal was cleaned by several cycles of Ar$^+$ bombardment and
annealing at $T$=900 K. The deposition rates were in the regime of
1~\AA\ per minute. The Cu(100) single crystal and the films were
characterized by means of low energy electron diffraction (LEED) and
Auger-electron spectroscopy (AES). The thickness of the ferromagnetic
films was calibrated by AES and by the signal-to-background ratio
(edge jump) at the respective $L_{2,3}$-edges. The thickness of the Fe
film was 50 monolayers (ML), and was 20 ML for the Co and Ni
films. The experimental data have been obtained accounting for
saturation effects, and the XMCD spectra have been corrected to
correspond to 100~\% circular polarization and collinear orientation
of the photon $\mathbf{k}$-vector and the magnetization. Details on
the sample preparation can be found in earlier
works.\cite{SWB04,Sche04}


\section{Results and discussion}   \label{sec-res}

\subsection{Magnetic moments}   \label{sec-moms}

\begin{table}
\caption{Magnetic moments (in units of \mB) for Fe, Co, and Ni
  calculated via the plain LSDA and via the \dmft\ with the AMF
  d.c.\ correction.}
\label{tab-mag}
\begin{ruledtabular}
\begin{tabular}{lddd}
  &  \multicolumn{1}{c}{Fe} & %
    \multicolumn{1}{c}{Co}  &  \multicolumn{1}{c}{Ni} \\
\hline
 \ms\ (LSDA)   &  2.26  &  1.60  &  0.63  \\
 \ms\ (\dmft)  &  2.18  &  1.65  &  0.68  \\ [0.5ex]
 \mo\ (LSDA)   &  0.052  &  0.079  &  0.051  \\
 \mo\ (\dmft)  &  0.094  &  0.155  &  0.071  \\
\end{tabular}
\end{ruledtabular}
\end{table}

Generally, if correlations are included via the DMFT, the spin
magnetic moment \ms\ of 3$d$ TMs changes only slightly while the
orbital magnetic moment \mo\ significantly increases.\cite{CMK08} This
is also illustrated by our results summarized in
Tab.~\ref{tab-mag}. We show results for the FLEX solver, the results
for the TMA solver differ by 3\% at most.  A detailed investigation of
the influence of the model parameters on magnetic moments has been
done by Chadov \ea\cite{CMK08}

\subsection{Shape of spectral peaks}   \label{sec-shape}

\begin{figure*}
\includegraphics[viewport=1.0cm 0.5cm 19.0cm 11.0cm]{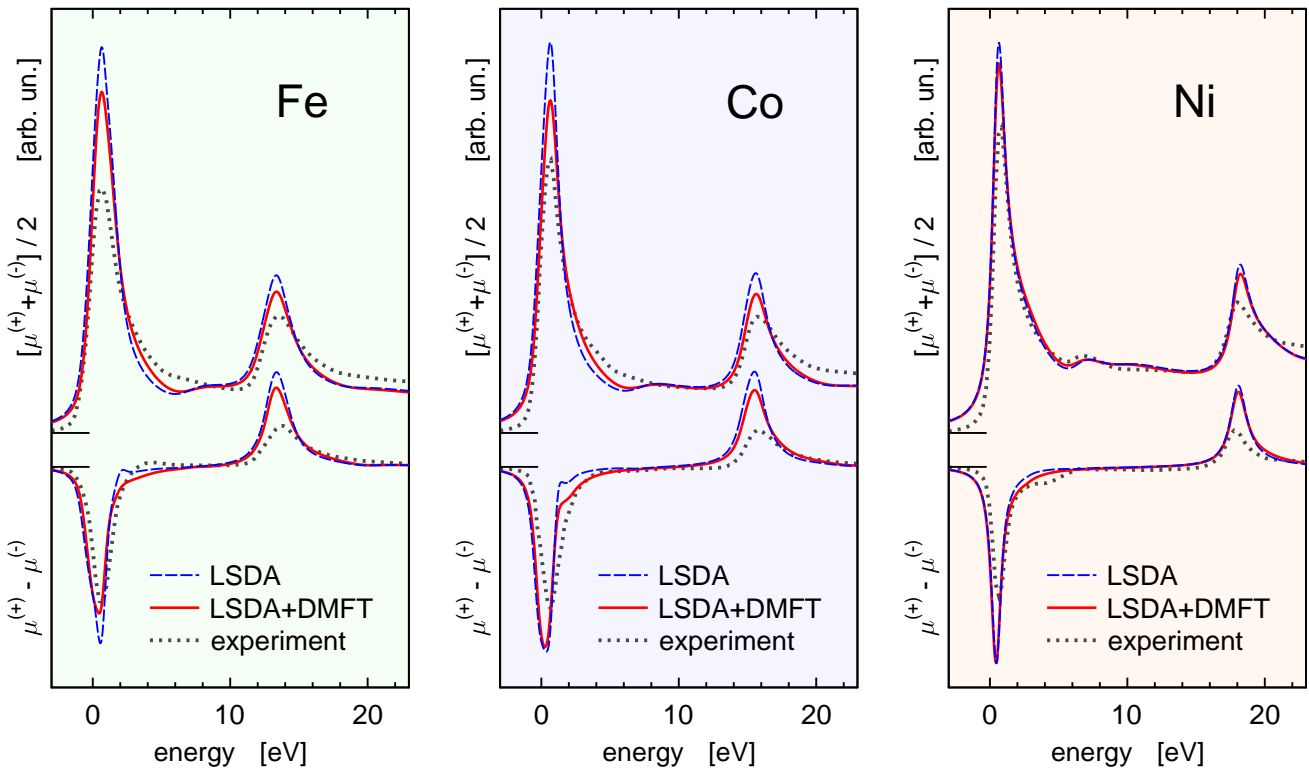}%
\caption{(Color online) \Led\ XANES and XMCD spectra of Fe, Co, and Ni
  calculated on the basis of the plain LSDA and of the \dmft\ with the
  AMF d.c.\ correction, compared to experiment.}
\label{fig-spectra}
\end{figure*}

\Led\ XANES and XMCD spectra calculated using the plain LSDA and using
the \dmft\ with the AMF d.c.\ correction are shown in
Fig.~\ref{fig-spectra}, together with experimental data. One can see
that including the correlations in the $d$ band changes the peak
intensities as well as the shapes of the main peaks.  In particular, a
more pronounced asymmetry of the $L_{3}$ peak appears in the
\dmft\ spectra, leading to a better agreement with experiment.  The
changes are, nevertheless, not very big.  This may appear surprising
given the fact that for photoemission spectra of these systems, the
inclusion of dynamic correlations via the \dmft\ has a very pronounced
effect as compared to the LSDA.\cite{BME+06} However, the self-energy
$\hat \Sigma(E)$\ is relatively small in the regime of unoccupied
states\cite{GDK+07} so it is actually plausible that its effect on the
XAS is not very pronounced.

\begin{figure*}
\includegraphics[viewport=1.0cm 0.5cm 19.0cm 11.0cm]{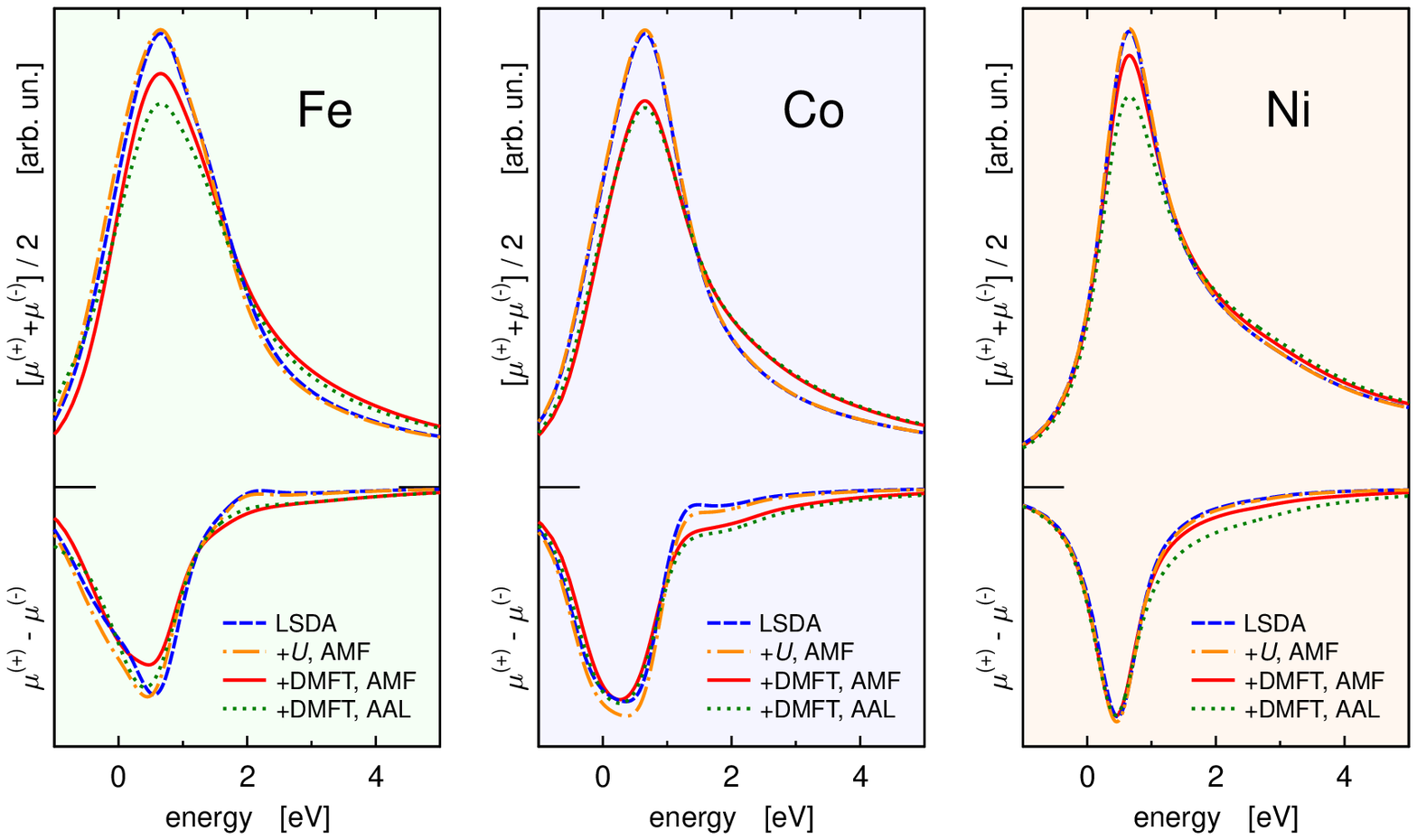}%
\caption{(Color online) $L_{3}$\ edge XANES and XMCD spectra for Fe,
  Co, and Ni calculated on the basis of the plain LSDA, via the
  \ldau\ with the AMF d.c.\ correction, via the \dmft\ with the AMF
  d.c.\ correction and via the \dmft\ with the AAL d.c.\ correction. }
\label{fig-flav}
\end{figure*}

As it was mentioned in the introduction, in order to remedy many
deficiencies of the LSDA as concerns the ground-state properties such
as \mo\ it is sufficient to include the correlations in a static way
only, via the OP Brooks scheme or via the \ldau.  However, these
schemes do not bring any significant improvement concerning XAS.  We
demonstrate this by showing in Fig.~\ref{fig-flav} spectra calculated
via the LSDA and via the \ldau\ (the dashed and dash-dotted lines).
Spectra obtained via the OP Brooks scheme are practically
indistinguishable from the \ldau\ results, so they are not shown here.
Fig.~\ref{fig-flav} shows that the \ldau\ method does not
significantly alter the calculated spectra of transition metals with
respect to the LSDA, similarly as it was found earlier for the OP
Brooks scheme.\cite{Ebe96c} This is in line with the concept that the
OP Brooks scheme can be seen in fact as one of the limits of the more
general \ldau\ concept.\cite{SLT98}

It was mentioned earlier in Sec.~\ref{sec-comput} that there are
several ways to correct for the d.c.\ error in the \dmft.  Even though
the AMF scheme seems to be the most reasonable d.c.\ procedure for
3$d$\ TMs, one cannot {\em a priori} exclude the possibility that
another d.c.\ scheme might be more appropriate for XAS.  Therefore, we
checked how the results change if the AAL d.c.\ scheme is employed
instead of the AMF scheme.  We found that the calculated spectra look
quite similar (Fig.~\ref{fig-flav}, dotted lines and full lines).  A
closer inspection of Fig.~\ref{fig-flav} reveals further that the
calculated spectra split into two groups, according to whether the
{\em dynamic effects} to the self-energy have been included
(\dmft\ calculations) or not (LSDA and \ldau\ calculations).  This is
especially evident for the XANES spectra; e.g., in the case of Co,
only two spectral curves can in fact be distinguished because the
results are pairwise practically identical (upper part of the middle
panel in Fig.~\ref{fig-flav}).  For the XMCD spectra, this splitting
of the four spectra into two groups is clearly visible at the
high-energy end of the Fe and Co $L_{3}$ peaks, between 1--3~eV in
Fig.~\ref{fig-flav}.  The choice of the d.c.\ model has thus only a
minor influence on the shapes of XANES and XMCD spectra.
Nevertheless, it has some influence on the intensities of the peaks
(see the following section).


\subsection{Relation to the \momsd\ ratio}   \label{sec-sumrul}

\begin{figure*}
\includegraphics[viewport=1.0cm 0.5cm 19.0cm 5.0cm]{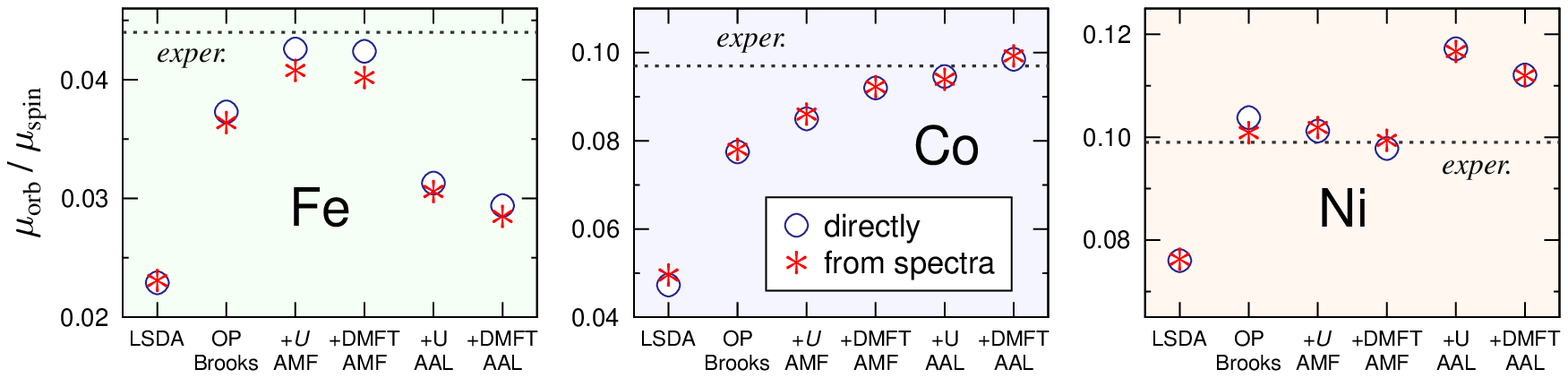}%
\caption{(Color online) The \momsd\ ratio evaluated directly and from
  theoretical XMCD spectra for calculations employing plain LSDA, OP
  Brooks, \ldau\ with the AMF d.c.\ correction (denoted as
  ``+$U$~AMF'' on the horizontal axis), \dmft\ with the AMF
  d.c.\ correction (``+DMFT~AMF''), \ldau\ with the AAL
  d.c.\ correction (``+$U$~AAL'') and \dmft\ with the AAL
  d.c.\ correction (``+DMFT~AAL''). The experimental ratios were taken
  those obtained from magneto-mechanical measurements.\cite{RF69}}
\label{fig-sumrul}
\end{figure*}

The changes in the XANES and XMCD spectra which result from the
inclusion the $d$\ band correlations are not very pronounced.  One may
ask to what degree are the changes significant at all: do they really
reflect the different treatment of the correlations?  The significance
of the changes in the spectra can be assessed by comparing them with
the changes of the ground-state properties.  A suitable quantity in
this respect is the \moms\ ratio, as it is underestimated by the plain
LSDA while it is correctly described by the \dmft\ method.  By
applying the XMCD sum rules (\ref{eq-spin})--(\ref{eq-orb}) to the
theoretical spectrum, the ratio between the $d$\ components of the
magnetic moments, \momsd, can be obtained and compared to the
\momsd\ ratio obtained directly from the ground-state electronic
structure.  If this is done for different ways of accounting for the
many body effects, the consistency of the changes in the spectra and
in the ground-state properties can be monitored.

Our results are summarized in Fig.~\ref{fig-sumrul}, where the
\momsd\ ratio evaluated by the two ways mentioned above is shown for
the plain LSDA, the OP Brooks scheme, the \ldau\ and the \dmft\ with
the AMF d.c.\ correction and the \ldau\ and the \dmft\ with the AAL
d.c.\ correction.  The \moms\ ratio derived from magneto-mechanical
experiments is shown for comparison.\cite{RF69} Use of \moms\ instead
of \momsd\ as an experimental reference is justified because both
ratios differ by less then 5\% and our focus is not on the agreement
of \mo\ with experiment (\mo\ depends also on the value of $U$
anyway).\cite{CMK08}

It is obvious that the changes of the intensities of the $L_{3}$\ and
$L_{2}$\ XMCD peaks reflect changes in \moms\ via the sum rules very
accurately.  The differences between spectra calculated by different
ways of dealing with the many body effects are thus relevant and
consistent.  Even relatively small changes in the intensities of XMCD
peaks in Fig.~\ref{fig-spectra} correspond to large changes in \mo, as
can be seen in Tab.~\ref{tab-mag}.  The failure of the LSDA to
reproduce the ratio of the $L_{3}$/$L_{2}$\ XMCD peak intensities thus
cannot be a simple consequence of the failure of the LSDA to yield
correct orbital moments \mo.


\subsection{Effect of the core hole within the final state approximation}   
\label{sec-corehole}

\begin{figure*}
\includegraphics[viewport=1.0cm 0.5cm 19.0cm 11.0cm]{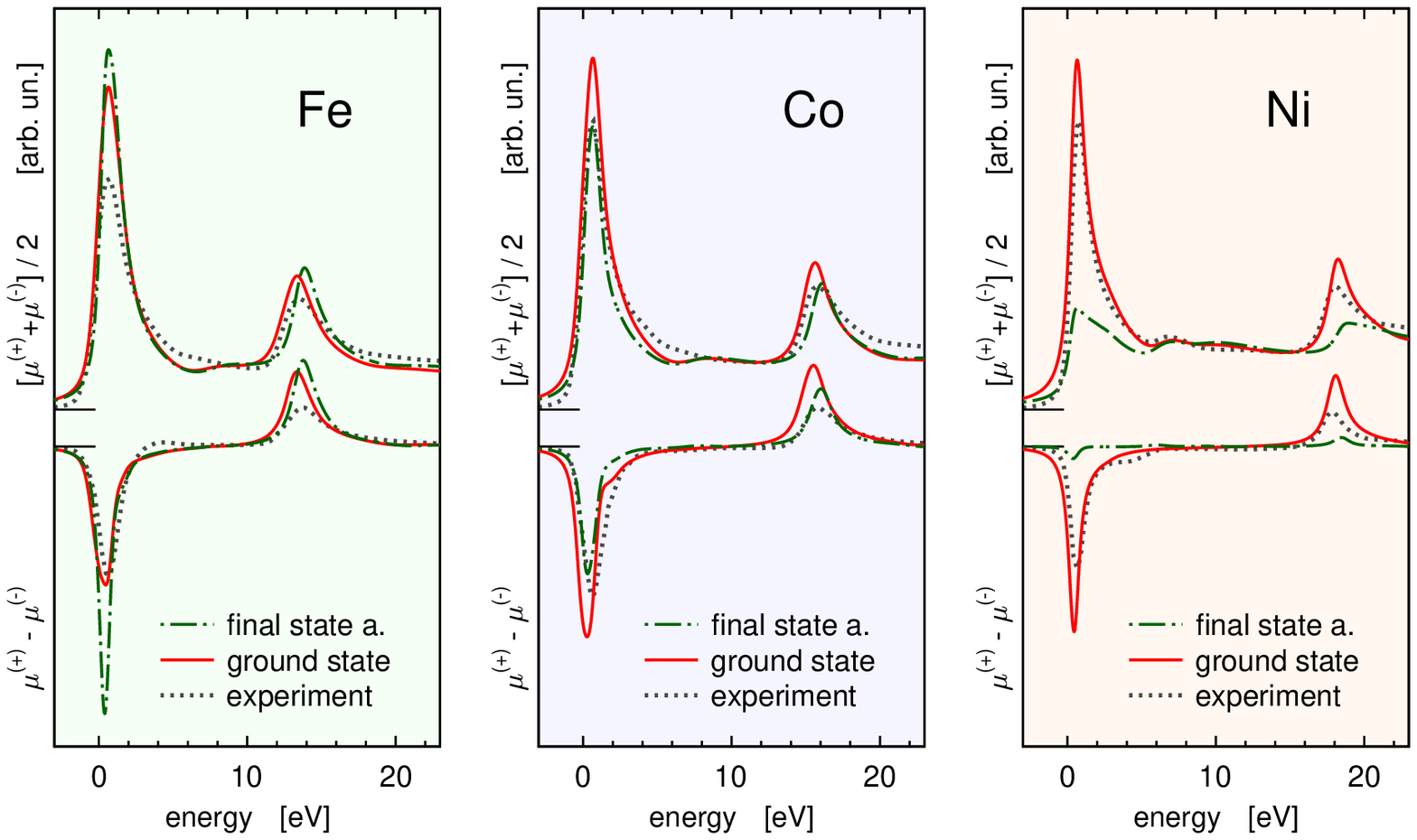}%
\caption{(Color online) \Led\ XANES and XMCD spectra of Fe, Co, and Ni
  calculated within the \dmft\ when the core hole is ignored (using
  the ground state potential) and when the core hole is included via
  the final state approximation. Experimental XANES and XMCD spectra
  are shown for comparison.}
\label{fig-core}
\end{figure*}

Accounting for the correlations in the $d$\ band via the
\dmft\ improves the calculated spectra but the improvement is not
dramatic (Fig.~\ref{fig-spectra}).  A better agreement with experiment
could presumably be obtained if the core hole was accounted for.  This
is quite a complicated task within the {\em ab initio} scheme.  A
technically relatively simple way of achieving this is via the
``static'' final state approximation (see Sec.\ \ref{sec-comput}).
For the $K$\ edges, such a scheme sometimes improves the XANES (e.g.,
for the Zn edge in ZnSe\cite{SMS97} or for the Si edge in
quartz)\cite{WJS98} while sometimes it has only a minor effect (early
transition metals).\cite{Zel88} However, the final state approximation
need not work for the \Leds\ which involve transitions to
semi-localized $d$\ states: by promoting a 2$p$\ electron into the
valence states one may effectively fill the $d$\ band of the
photoabsorbing atom, suppressing to a large extent the intensity of
the white line.  This issue will be explored in the following.

To see the effect of the final state approximation, we applied it on
top of the \dmft\ procedure.  The results are shown in
Fig.~\ref{fig-core}.  It is evident that this procedure does not
generally improve the \Led\ spectra of late 3$d$\ TMs.  Hardly any
systematic trend in the effect of the core hole can be found ---
neither as concerns the shape of the white lines, nor as concerns the
ratio of the intensities of the $L_{3}$\ and $L_{2}$\ peaks.  For Fe,
we observe an increase of the ratio of intensities of the $L_{3}$\ and
$L_{2}$\ XMCD peaks, in agreement with experiment.  For Co, the final
state approximation produces only minor changes with respect to the
ground state calculations.  For Ni, however, including the core hole
via the final state approximation substantially worsens the agreement
with experiment --- the XAS white lines as well as the prominent XMCD
peaks practically disappear!  A similar situation occurs if the final
state approximation is applied over the plain LSDA (therefore, only
the \dmft\ results are shown here).

We tested this scheme also by using a half-filled core hole, i.e.,
employing the Slater transition state method.  Sometimes this
procedure works well for the $K$\ edge spectra.\cite{MT00,LLL10}
Satisfying results were also reported for the Cu $L_{3}$\ edge
XANES.\cite{LMH01} In our case, however, the Slater transition state
method does not bring any substantial improvement --- the results just
lie approximately half-way between the results obtained without a core
hole and with a full core hole, so they are not shown here.

Our results demonstrate that the final state approximation is
unsuitable for describing the \Led\ XAS in late 3$d$\ TMs.  This
applies for calculations based on the plain LSDA as well as on the
\dmft\ method.


\section{Conclusions}   \label{sec-zaver}

Our goal was to find out whether the differences commonly occurring
between the experimental \Led\ XANES and XMCD spectra of
3$d$\ transition metals on the one hand and {\em ab-initio}
calculations on the other hand are mainly due to the way the LSDA
deals with the correlations.  By performing the \dmft\ calculations,
we found that if valence- and conduction-band correlations are
included, the spectra change in the right direction with respect to
the plain LSDA.  In particular, the \dmft\ yields asymmetric $L_{3}$
XAS white lines.  The improvement is, however, rather incremental than
dramatic.  The ratio of the intensities of the $L_{3}$\ and
$L_{2}$\ XAS and XMCD peaks is not significantly improved by the
\dmft\ formalism.  The changes in the intensities of the XMCD peaks
are, nevertheless, consistent with the changes of the \moms\ ratio as
suggested by the XMCD sum rules.

When the core hole is additionally accounted for within the final
state approximation, the agreement between theory and experiment
generally does not improve (in the case of Ni, it substantially
worsens).  It appears, therefore, that to get a decisive improvement
of {\em ab-initio} calculations of the \Led\ XAS and XMCD spectra of
3$d$\ metals, the dynamic correlations between the excited
photoelectron and the core hole have to be taken into account.


\begin{acknowledgments}
This work was supported by the Grant Agency of the Czech Republic
within the project 108/11/0853, by the Deutsche Forschungsgemeinschaft
through FOR 1346 and by the German ministry BMBF under contract
05KS10WMA.  The research in the Institute of Physics AS~CR was
supported by the Academy of Sciences of the Czech Republic.
\end{acknowledgments}





%

\end{document}